\documentclass[a4paper,11pt]{article}
\usepackage{amssymb}
\usepackage{amsmath}
\usepackage{hyperref}
\usepackage{url}
\usepackage{stmaryrd}
\usepackage[numbers,square,sort&compress]{natbib}
\usepackage{graphicx}

\usepackage{subcaption}

\usepackage{tikz}
\usetikzlibrary{arrows}
\usetikzlibrary{calc, decorations.pathmorphing}

\usepackage{xspace}
\usepackage{color}
\usepackage{longtable}
\usepackage{listings}
\DeclareGraphicsExtensions{.pdf,.png,.jpg}

\usepackage{commath}
\usepackage{stmaryrd}

\usepackage[margin=2.5cm]{geometry}

\newcommand\Ray{\mathit{Ra}}

\begin{document}
\title{\bfseries CO$_2$ dissolution  in a background hydrological flow}
%
\author{H.~Juliette~T.~Unwin\thanks{Department of Engineering,
    University of Cambridge (\url{hjtu2@cam.ac.uk})}
  \and
  Garth~N.~Wells\thanks{Department of Engineering, University of
    Cambridge (\url{gnw20@cam.ac.uk})}
  \and
  Andrew~W.~Woods\thanks{BP Institute, University of Cambridge
    (\url{andy@bpi.cam.ac.uk})}}
\date{}
\maketitle
\begin{abstract}

During CO$_2$ sequestration into a deep saline aquifer of finite
vertical extent, CO$_2$ will tend to accumulate in structural highs
such as offered by an anticline. Over times of tens to thousands of
years, some of the CO$_2$ will dissolve into the underlying
groundwater to produce a region of relatively dense, saturated water
directly below the plume of~CO$_2$.  Continued dissolution then
requires the supply of unsaturated aquifer water.  In an aquifer of
finite vertical extent, this may be provided by a background
hydrological flow, or a laterally-spreading buoyancy-driven flow
caused by the greater density of the CO$_2$ saturated water relative
to the original aquifer water.

We investigate the long time steady-state dissolution in the presence
of a background hydrological flow.  In steady-state, the distribution
of CO$_2$ in the groundwater upstream of the aquifer involves a
balance between three competing effects: (i) the buoyancy-driven flow
of CO$_2$ saturated water; (ii) the diffusion of CO$_2$ from saturated
to under-saturated water; and (iii) the advection associated with the
oncoming background flow.  This leads to three limiting regimes.  In
the limit of very slow diffusion, a nearly static intrusion of dense
fluid may extend a finite distance upstream, balanced by the pressure
gradient associated with the oncoming background flow.  In the limit
of fast diffusion relative to the flow, a gradient zone may become
established in which the along aquifer diffusive flux balances the
advection associated with the background flow. However, if the
buoyancy-driven flow speed exceeds the background hydrological flow
speed, then a third, intermediate regime may become established.  In
this regime, a convective recirculation develops upstream of the
anticline involving the vertical diffusion of CO$_2$ from an upstream
propagating flow of dense CO$_2$ saturated water into the downstream
propagating flow of CO$_2$ unsaturated water.  For each limiting case,
we find analytical solutions for the distribution of CO$_2$ upstream
of the anticline, and test our analysis with full numerical simulations.
A key result is that, although there may be very different controls on
the distribution and extent of CO$_2$ bearing water upstream of the
anticline, in each case the dissolution rate is given by the product
of the background volume flux and the difference in concentration
between the CO$_2$ saturated water and the original aquifer water
upstream.
\end{abstract}

\section{Introduction}
\label{sec:introduction}

\begin{figure}
  \begin{subfigure}[b]{\textwidth}
    \center\includegraphics[width=0.75\textwidth]{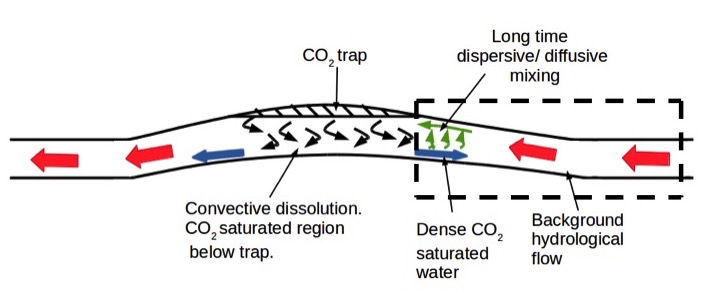}
    \caption{Cartoon of the geological problem with region of interest
      indicated by dashed box. The large red arrows represent the
      direction of the background hydrological flow, while the smaller
      black curved arrows below the CO$_2$, which is trapped at the
      top of the anticline, represent the convective mixing of the
      CO$_2$ saturated and undersaturated water. The smaller blue
      arrow inside the dashed box represents the buoyancy driven flow
      of dense CO$_2$ saturated water flowing upstream into the
      background hydrological flow.}
    \label{fig:cartoon}
  \end{subfigure}
  \begin{subfigure}[b]{\textwidth}
    \vspace{1.5em}
    \centering
    \begin{tikzpicture}[scale=16]
      \draw[very thick] (0,0) -- (0, .1)  node [midway, anchor=east]
    	     {$\Gamma_{c}$};
   	\draw[very thick] (0.5,0) -- (0.5, .1)  node [midway, anchor=west]
    	{$\Gamma_{B}$};
    	\draw[very thick] (0,0) -- (0.5, 0)  node [midway, anchor=north]
    	{$\Gamma_{\text{cap}}$};
    	\draw[very thick] (0,.1) -- (0.5, .1)  node [midway, anchor=south]
    	{$\Gamma_{\text{cap}}$};
     	\draw[solid, <->] (0.54,0) -- (0.54, .1)  node
      	[midway,anchor=west] {$H$};
      	\draw[solid, <->] (0,-0.033) -- (0.5, -0.033)  node
      	[below,midway,anchor=north] {$L$};
     	\draw[solid, ->] (-0.031,-0.05) -- (-0.031, 0.002)  node
     	[above,anchor=south]{$y$};
     	\draw[solid, ->] (-0.031,-0.05) -- (0.01, -0.05)
     	node [above,anchor=west]{$x$};
     	\draw[ultra thick, <-] (0.45,0.015) -- (0.51, 0.015) ;
     	\draw[ultra thick, <-] (0.45,0.035) -- (0.51, 0.035) ;
     	\draw[ultra thick, <-] (0.45,0.065) -- (0.51, 0.065) ;
     	\draw[ultra thick, <-] (0.45,0.085) -- (0.51, 0.085) ;
     	\draw[ultra thick, ->] (-0.02,0.015) -- (0.05, 0.015) ;
     	\draw[ultra thick, ->] (0.05,0.085) -- (-0.02, 0.085) ;
     	\node[text width=.1cm] at (0.44,0.05){$u_{B}, c_{0}$};
     	\node[text width=.1cm] at (0.06,0.015){$c_{D}$};
     	\node[text width=.1cm] at (0.25,0.05){$c_{0}$};
    \end{tikzpicture}
    \caption{Model problem for analysis and simulation.}
    \label{fig:boundaries}
  \end{subfigure}
  \caption{Problem of interest.}
  \label{fig:problem}
\end{figure}
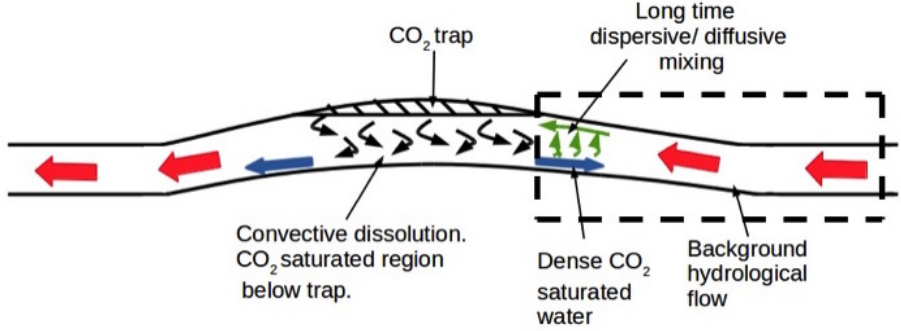
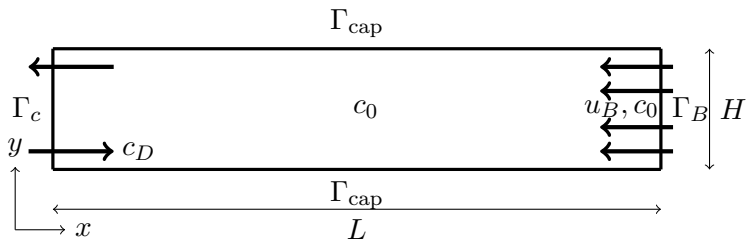

Carbon capture and storage in deep saline aquifers has been proposed
as a potential means to limit carbon emissions into the atmosphere,
while enabling the continued supply of energy from fossil fuels. Much
research has been undertaken to explore the processes which control
the storage of CO$_2$ over very long periods, and in particular the
integrity of a geological storage facility in terms of the possible
migration of CO$_2$ back to the surface \citep{hesse:2007, boait:2012,
  verdon:2013, pruess:2003}. Owing to the buoyancy of CO$_2$ relative
to water at depths of 1--2~km, CO$_2$ tends to migrate along permeable
sedimentary layers and ultimately ponds in structural highs, for
example an anticline, which represents the upper part of a fold or
other deformation in the geological strata (see
Figure~\ref{fig:cartoon}).  Such structural traps offer a possible
storage site providing there is a competent seal rock above the
anticline (e.g.,~\citealt{IPCC_2005}). However, CO$_2$ is soluble in
groundwater, which may accommodate concentrations of a few wt\% CO$_2$
in solution.  This in turn leads to an increase in density of the
water. With the dependency of water density on CO$_{2}$ concentration,
convectively-driven dissolution may develop. Water below the trapped
CO$_2$ plume becomes increasingly concentrated in CO$_2$ until it
becomes convectively unstable and sinks into the underlying permeable
rock, to be replaced by less dense, unsaturated water
(c.f.~\citealt{riaz:2006, pau:2009, hewitt:2014, lindeberg:1997}).
Eventually, the water below the CO$_2$ plume becomes fully saturated
and the continued dissolution requires a more distal supply of
undersaturated groundwater.

\citet{szulczewski_carbon_2013} examined the longer-time dissolution
by examining the convective exchange flow which can develop in a
horizontal aquifer. They established that following the initial
dissolution and near-saturation of the groundwater directly below the
plume of CO$_2$, the lateral convective exchange flow leads to the
slow horizontal spreading of a zone of CO$_2$ enriched groundwater
associated with the continued dissolution. Eventually, the dynamics of
this zone may become controlled by a balance between: (a)~the
buoyancy-driven shear, as the dense groundwater spreads along the base
of the aquifer; and (b) the vertical diffusion of CO$_2$ from this
outward spreading dense fluid to the return flow of under-saturated
groundwater higher in the aquifer. By itself, such buoyancy-driven
shear dispersion leads to a progressively waning rate of dissolution,
and, owing to the relatively low solubility of CO$_2$ in the ground
water, the prediction that the plume of CO$_2$ may be trapped in the
anticline for a very long time \citep{szulczewski_carbon_2013}.

However, at long times the slow background hydrological flows which
transport fluid laterally through aquifers will become important in
controlling the flux of unsaturated water from far upstream,
especially as other transport processes wane.  It is the purpose of
this paper to explore the long term influence of a background
hydrological flow on the process.  In this context, \citet{woods:2012}
established some non-linear bounds on the flux of groundwater that may
reach an anticline along a weakly tilted aquifer resulting from the
interaction of a background hydrological flow with a convective
exchange flow for intermediate times, during which the cross-aquifer
diffusive transport of CO$_2$ is small. In the present work, we
account for the effects of such diffusion and this leads to a more
complex problem involving the interaction of the background advection,
the buoyancy-driven flow and the diffusive transport of CO$_2$
upstream of the anticline.  We note that in our modelling we assume
the background hydrological flows are constant in time and that there
are no mineralogical reactions of the CO$_2$ with the formation; these
are simplifications but provide a reference with which the effects of
mineral precipitation or changes in the background forcing over time
could be compared.

We develop a series of idealised, analytical solutions for the
governing equations and then test these solutions using a full
numerical simulation of the two-dimensional governing equations.  In
modelling the flow in porous rock, we assume that the dynamics are
governed by Darcy's Law, which relates to slow viscous flow.  This is
an appropriate model in the present content of slow hydrological flows
and the slow buoyancy driven flow of CO$_2$ saturated water
(e.g.~\citealt{woods:2015, bear:1972}). We thereby establish that when
the buoyancy-driven flow of the dense CO$_2$ saturated water is large
compared to the background flow speed, which is typical, three
different regimes may become established: (i)~the weak diffusion limit
in which there is a nearly static intrusion of CO$_2$ saturated water
upstream; (ii)~an intermediate regime in which there is a balance of
buoyancy-driven flow and vertical diffusion with the oncoming flow;
and (iii)~a strong diffusion limit, in which there is a balance
between the upstream diffusion of CO$_2$ and the downstream advective
transport of unsaturated water.  In each case, the dissolution rate is
proportional to the groundwater flow, even though the controls on the
extent of the CO$_2$ enrichment of the groundwater upstream of the
anticline may be very different.

\section{Model system}
\label{sec:full_model}

In our analysis, we consider a two-dimensional flow geometry, shown in
Figure~\ref{fig:problem}. This corresponds to an anticline produced by
a fold in the geological strata that extends for a relatively long
distance in the direction normal to the page compared to the width of
the fold.  We consider a background hydrological flow that supplies
fluid from the right-hand side, and an aquifer that is of uniform
thickness and horizontal.  We assume that directly below the plume of
CO$_2$ which is trapped in the anticline, the water is fully saturated
in CO$_2$ as a result of the vertical convective dissolution, and that
this is carried downstream (to the left in
Figure~\ref{fig:cartoon}). The primary purpose of this paper is to
examine how the concentration of CO$_2$ varies in the upstream
direction, shown in Figure~\ref{fig:boundaries}.  In our numerical
model, we choose the location of the upstream boundary of the flow
domain to be upstream of the region containing elevated concentrations
of CO$_2$, so that we can
impose a simple uniform flow of fluid with uniform background CO$_2$
concentration.

The full model involves Darcy flow with a buoyancy term that is
dependent on the dissolved CO$_{2}$ concentration~$c \in [c_{0},
c_{D}]$, where $c_{0} \ge 0$ is the initial CO$_2$ mass fraction of
the groundwater and $c_{D}$ is the mass fraction of the CO$_2$
saturated groundwater below the trapped plume of CO$_{2}$ (see
Figure~\ref{fig:boundaries}). We assume a fluid density $\rho$ given
by
\begin{equation}
  \rho = \rho_{0} +  \beta (c - c_{0}) \rho_{0},
\end{equation}
where the constant $\rho_{0} \ge 0$ is the initial water density and
$\beta \ge 0$ is the expansion coefficient of dissolved CO$_2$ in
groundwater.  We work with scaled concentration $c^{\star} \in [0,1]$,
given by
\begin{equation}
  c = (c_{D} - c_{0}) c^{\star} + c_{0}.
 \end{equation}

To formulate the governing equations in non-dimensional form, we
denote dimensionless variables by the superscript~`$\star$' and we
introduce:
\begin{align}
  \boldsymbol{u} & = \frac{\kappa\beta(c_{D}-c_{0})\rho_{0}g}
  {\mu}\boldsymbol{u}^{\star},
  \\
  p & = \beta(c_{D}-c_{0}) \rho_{0} g H p^{\star},
  \\
  t & = \frac{H \mu }{\kappa\beta(c_{D}-c_{0})\rho_{0}g}t^{\star},
  \\
  \boldsymbol{x} &=  H\boldsymbol{x}^{\star},
\end{align}
where $\boldsymbol{u}$ is the Darcy velocity, the constants $\kappa
\ge 0$ and $\mu > 0$ are the permeability and viscosity, respectively,
$g$ is the gravitational acceleration, $p$ is the pressure field, $H$
is the characteristic height of the domain, $t$ is time and
$\boldsymbol{x}$ is spatial position.

We denote our domain of interest by $\Omega \subset \mathbb{R}^{2}$,
with boundary $\Gamma = \partial\Omega$ and outward unit normal vector
to the boundary~$\boldsymbol{n}$.  The boundary is partitioned as
depicted in Figure~\ref{fig:boundaries}.  We formulate a
time-dependent model, with time interval of interest denoted by~$I =
[0, t_N)$. We are interested in steady solutions, hence $t_{N}$ will
be chosen to be suitably large in numerical simulations.  In terms
of non-dimensional quantities, the continuity equation, the Darcy equation
and the boundary conditions read:
\begin{alignat}{2}
  \label{eq:contmassnd}
  \nabla^{\star} \cdot \boldsymbol{u}^{\star} &= 0
  &\quad \textrm{on} \ \Omega \times I,
  \\
  \label{eq:darcynd}
  \boldsymbol{u}^{\star} &= - \nabla^{\star} p^{\star} + c^{\star}
  \boldsymbol{e}_{k}
  &\quad \textrm{on} \ \Omega \times I,
  \\
  \label{eq:pdbcnd}
  p^{\star} &= p^{\star}_{D}
  &\quad \textrm{on} \ \Gamma_{c} \times I,
  \\
  \label{eq:pnbcnd2}
  \boldsymbol{u}^{\star} \cdot \boldsymbol{n} &= 0
  &\quad \textrm{on} \ \Gamma_{\text{cap}} \times I,
  \\
  \label{eq:pnbcnd}
  \boldsymbol{u}^{\star} \cdot \boldsymbol{n} &= - u_{B}^{\star}
  &\quad \textrm{on} \ \Gamma_{B} \times I,
\end{alignat}
where $\boldsymbol{e}_{k}$ is the unit vector in the direction in
which gravity acts, $p^{\star}_{D}$ is a prescribed pressure and
$u_{B}^{\star} \geq 0$ is the prescribed fluid velocity across the
inflow boundary.  The condition in~\eqref{eq:pnbcnd} gives a
background flow from right-to-left in Figure~\ref{fig:boundaries}.

The concentration of dissolved CO$_2$ is modelled by:
\begin{alignat}{2}
  \label{eq:advdiffnd}
  \frac{\partial c^{\star}}{\partial t^{\star}}
  + \nabla^{\star} c^{\star} \cdot \boldsymbol{u}^{\star}
  - \frac{1}{\Ray} \nabla^{\star} \cdot \nabla^{\star} c^{\star}
     &= 0  &\quad \textrm{on} \ \Omega\times I,
  \\
  \label{eq:cbc1And}
  c^{\star} \boldsymbol{u}^{\star} \cdot \boldsymbol{n}
  &= \boldsymbol{u^{\star}} \cdot \boldsymbol{n}
  &\quad \textrm{on} \ \Gamma_{c, {\rm in}} \times I,
  \\
  \label{eq:dbc1and}
  \frac{1}{\Ray}\nabla^{\star} c^{\star}\cdot\boldsymbol{n} &= 0
  &\quad \textrm{on} \ \Gamma_{c} \times I,
  \\
  \label{eq:cdbc1bnda}
  (-\frac{1}{\Ray}\nabla^{\star} c
   + c^{\star}\boldsymbol{u}^{\star}) \cdot \boldsymbol{n} &= 0
  &\quad\textrm{on} \ \del{\Gamma_{\text{cap}} \cup \Gamma_{B}} \times I,
  \\
  c^{\star}(x,0) &= 0 &\quad \textrm{on} \ \Omega,
\end{alignat}
where $\Gamma_{c, {\textrm{in}}}$ is the portion of $\Gamma_{c}$ on which
$\boldsymbol{u}^{\star} \cdot \boldsymbol{n} < 0$, the constant
$\Ray$ is a Rayleigh number,
\begin{equation}
  \Ray = \frac{\kappa\beta(c_{D}-c_{0})\rho_{0} gH}{\mu D},
\end{equation}
and $D \ge 0$ is the pore-scale diffusivity.  The boundary condition
in~\eqref{eq:cbc1And} ensures that the advective flux of CO$_2$ at
the CO$_2$ trap boundary ($x=0$) has dimensionless concentration
unity on the inflow parts of the
boundary, while it is not prescribed on the outflow parts of the
boundary.  In steady state,
equations~\eqref{eq:advdiffnd}--\eqref{eq:cdbc1bnda} require that:
\begin{equation}
  \int_{\Gamma_c} c^{\star} \boldsymbol{u}^{\star} \cdot
  \boldsymbol{n} \dif s = 0
\end{equation}
We work from this point onward with the non-dimensional equations,
hence we drop the~`${\star}$' superscript in the following.

\section{Physical discussion}
\label{sec:defeq}

The above non-dimensionalisation identifies two controlling
parameters: $u_{B}$ represents the ratio of the background flow speed
to the buoyancy-driven flow speed, and $\Ray$ represents the
buoyancy-driven flow speed compared to the effective speed associated
with vertical diffusive transport across the flow domain.  These two
parameters may be used to delineate the different flow regimes which
may develop. We explore this below.

\subsection{Gravity intrusion model}
\label{sec:gravity_intrusion}
In the case of weak diffusion, we expect that a nearly static
intrusion of the dense CO$_2$ saturated fluid extends upstream into
the aquifer, and that this is balanced by the pressure gradient of the
oncoming hydrostatic flow. There will be a thin diffusive boundary
layer between the intrusion of CO$_2$ saturated water and the oncoming
flow of groundwater. If the intrusion extends far into the aquifer ($X
\gg 1$), then the continuity equation suggests that the background
flow will be largely parallel to the boundary of the domain.  To model
this regime, we assume a sharp interface in the concentration field at
a height $h(x)$  above the lower boundary of the aquifer, where
$0 \le h(x) \le 1$. This interface delineates the CO$_2$ saturated
intrusion and the overlying
groundwater.  Assuming the pressure in the intrusion is approximately
hydrostatic (c.f.~\citet{woods:2015, huppert:1995}), then in
equilibrium the buoyancy driven pressure gradient in the $x$-direction
along the intrusion matches the pressure gradient associated with the
background flow above the intrusion, which has speed $ u_{B}/(1 -
h)$. This leads to the balance
\begin{equation}
  -(1 - h) \od{h}{x} = u_{B}.
\end{equation}
The shape of the intrusion is therefore given by
\begin{equation}
  h(x) = 1 - \sqrt{2u_{B}x},
  \label{eq:intrusion_height}
\end{equation}
and it follows that the extent of the intrusion into the aquifer is
$X_{\rm int}= 1/2u_{B}$.  This implies that if $u_{B}$ is small, the
intrusion extends far upstream into the aquifer, relative to the
vertical extent of the aquifer, and the assumption
that the flow is one-dimensional is valid.

The dimensionless time-of-travel of the oncoming flow past this
intrusion is given by $\tau_{\rm int} = \int_{X_{\rm int}}^{0}
(1/u) \dif x = 1/(3 u^{2}_{B})$.  For the interface to remain sharp,
the time should be small relative
to the diffusion time, $\tau_{\rm diff} = \Ray$. This requires that
$\Ray \gg 1/(3u_{B}^{2})$ which may be expressed in the form:
\begin{equation}
  u_{B}^{2} \Ray \gg \frac{1}{3}.
  \label{eq:intrusion}
\end{equation}
For simplicity, we will henceforth use the condition~$u_{B}^{2} \Ray
\gg 1$.

When $u_{B}$ is large, we expect any intrusion will become progressively
smaller, with $X_{\rm int} \le 1/u_{B}$ (for example
Figure~\ref{fig:flowplots2-intrusion}, Section~\ref{sec:stong_bg_flow})
and so we now expect the time-scale $1/u^{2}_{B}$ to be an upper bound on
the advection time, $\sim X_{\rm int}/u_{B}$. Since $X_{\rm int}<1$ for
large $u_{B}$, we compare this with the diffusion time along, rather
than across, the aquifer. This diffusion time scales as $X^2_{\rm int}
\Ray$ and suggests that the line $\Ray =1$ provides an upper bound on
the transition from the diffusion to the advection regime. We return to
this case in Section~\ref{sec:stong_bg_flow}.

\subsection{Buoyancy-driven shear dispersion model}

In the case $u_{B}^{2} \Ray \ll 1$, diffusion in the vertical
direction will be relatively fast, hence the vertical gradient in
concentration across the aquifer will be small. However, there may be
a significant gradient in the along-aquifer direction.  This can lead
to different regimes in which diffusion is important and we now
establish conditions which determine whether a buoyancy-driven shear
flow develops or a simple advection--diffusion balance controls the
transport.  We explore these two limits by starting from the full
equations and allowing for variations in the concentration of CO$_2$
in the fluid associated with the diffusive flux. We follow largely the
analysis of \citet{szulczewski_carbon_2013} and \citet{woods:2015} to
formulate a one-dimensional asymptotic model for the long-time
evolution of the vertically averaged concentration field, but now in
the presence of a background flow.

We decompose the CO$_2$ concentration of the groundwater in the form
$c(x, y, t) = \bar{c}(x, t) + \hat{c}(x, y, t)$, where $\bar{c}$
is the average concentration across the depth of the aquifer:
\begin{equation}
  \bar{c} =  \int^{1}_{0} c \dif y.
\end{equation}
Under the assumptions that the concentration fluctuations $\hat{c}$
are small, as expected in the limit $u_{B}^{2} \Ray \ll 1$, and that
the horizontal scale of the flow is much larger than the thickness of
the aquifer, as expected in the case $u_{B} \ll 1$, the
non-hydrostatic vertical pressure gradient is relatively small and the
flow is approximately parallel to the boundaries of the flow domain.
Therefore the pressure may be approximated by:
\begin{equation}
  p = p_0 - y\Bar{c},
  \label{eq:pressure}
\end{equation}
where $p_{0} = p_{0}(x, t)$ is the pressure at the base of the aquifer.

We decompose the velocity of the fluid $\boldsymbol{u} = (u, v)$ into
a sum of the average across the depth of the aquifer
$\Bar{\boldsymbol{u}} = (\Bar{u}, \Bar{v})$ and the fluctuation
$\Hat{\boldsymbol{u}} = (\Hat{u}, \Hat{v})$, where:
\begin{equation}
  \Bar{\boldsymbol{u}} = \int^{1}_{0} \boldsymbol{u} \dif y.
\end{equation}
Using the approximation for the pressure~\eqref{eq:pressure}, Darcy's
law implies that
\begin{equation}
  \label{eq:uplusp}
  u = -\frac{\partial p_{0}}{\partial x}
  + y\frac{\partial{\bar{c}}}{\partial x},
\end{equation}
and so
\begin{equation}
  \label{eq:uflucprof}
  \hat{u}
  = \frac{\partial \bar{c}}{\partial x}\del{y - \frac{1}{2}}.
\end{equation}

Taking the vertical average of the transport
equation~\eqref{eq:advdiffnd}, and combining with the continuity
equation~\eqref{eq:contmassnd}, it may be shown that
\begin{equation}
  \label{eq:transportbuoyancy}
  \frac{\partial \bar{c}}{\partial t}
  + \bar{u}\frac{\partial \bar{c}}{\partial x}
  + \overline{\hat{u}\frac{\partial\hat{c}}{\partial x}}
  + \overline{\frac{\partial\hat{u}}{\partial x}\hat{c}}
  = \frac{1}{\Ray}\frac{\partial^{2}\bar{c}}{\partial x^{2}}.
\end{equation}
Subtracting~\eqref{eq:transportbuoyancy} from the transport equation,
we obtain an equation governing the evolution of the concentration
fluctuation:
\begin{equation}
  \frac{\partial \hat{c}}{\partial t}
  + \Hat{u}\frac{\partial{\bar{c}}}{\partial{x}}
  + \Bar{u}\frac{\partial \Hat{c}}{\partial x}
  + \Hat{u}\frac{\partial \Hat{c}}{\partial x}
  + \Hat{v}\frac{\partial \Hat{c}}{\partial y}
  = \frac{1}{\Ray}\del{ \frac{\partial^{2}\hat{c}}{\partial x^2}
    + \frac{\partial^{2}\hat{c}}{\partial y^2}}
  + \overline{\hat{u}\frac{\partial \hat{c}}{\partial x}}
  +\overline{\frac{\partial\hat{u}}{\partial x}\hat{c}}.
  \label{eq:fluctuationsinc}
\end{equation}

After long time periods, we expect the dominant balance in
equation~\eqref{eq:fluctuationsinc} to be between the distortion of
the mean concentration due to the shear flow and the cross layer
diffusion~\citep{taylor:1953}.  This gives rise to the following
dominant balance, which can be shown \emph{a posteriori}:
\begin{equation}
  \label{eq:balance}
  \frac{1}{\Ray}\frac{\partial^{2}\hat{c}}{\partial y^{2}}
  = \hat{u}\frac{\partial \bar{c}}{\partial x}.
\end{equation}
Inserting~\eqref{eq:uflucprof} into~\eqref{eq:balance} and integrating
leads to the expression
\begin{equation}
  \hat{c} = \Ray \del{\frac{\partial \bar{c}}{\partial x}}^{2}
  \del{\frac{y^3}{6} - \frac{y^2}{4} + \frac{1}{24}}.
\end{equation}
Combining this expression with the expression for $\Hat{u}$
in~\eqref{eq:uflucprof}, the depth-averaged transport
equation~\eqref{eq:transportbuoyancy} becomes:
\begin{equation}
  \label{eq:transportbuoyancyfull}
  \frac{\partial \bar{c}}{\partial t}
  - u_{B}\frac{\partial\bar{c}}{\partial x}
  = \frac{1}{\Ray}\frac{\partial^{2}\bar{c}}{\partial x^{2}}
  + \frac{\Ray}{120}\frac{\partial}{\partial x}
  \del{\frac{\partial \bar{c}}{\partial x}}^{3}.
\end{equation}

At long times,~\eqref{eq:transportbuoyancyfull} admits steady
solutions in which $\bar{c} \rightarrow 0$ as $x \rightarrow
\infty$. In the limit that $u^2_B \Ray \ll 1$, the vertical gradient
of concentration is small, and so in this limit it follows from the
boundary condition in~\eqref{eq:cbc1And} that these solutions also
require $\bar{c} \approx 1$ at $x = 0$.  To help interpret these
solutions, it is convenient to re-scale the horizontal coordinate
according to
\begin{align}
  x = \del{\frac{\Ray}{120 u_{B}}}^{\frac{1}{3}} \Tilde{x},
\end{align}
leading to the relation
\begin{equation}
  \label{eq:simplificationalpha}
  -\frac{\partial \bar{c}}{\partial \Tilde{x}}
  = \alpha\frac{\partial^{2}\bar{c}}{\partial \Tilde{x}^{2}}
  + \frac{\partial}{\partial \Tilde{x}}
  \del{\frac{\partial\bar{c}}{\partial \Tilde{x}}}^{3},
\end{equation}
where
\begin{align}
  \label{eq:alpha}
  \alpha = \del{\frac{120}{\Ray^{4}u_{B}^{2}}}^{\frac{1}{3}}.
\end{align}
We see that for large $\alpha$ diffusion dominates ($-\Bar{c} = \alpha
\Bar{c}^{\prime} $), and for small $\alpha$ dispersion is dominant
($-\Bar{c} = \Bar{c}^{\prime 3} $).
We have not found an analytic solution
to~\eqref{eq:simplificationalpha}, but in the two limits $\alpha \gg
1$ and $\alpha \ll 1$ there are useful analytical approximations.

In the limit $\alpha \ll 1 $, the buoyancy-driven dispersion balances
the advection. The solution to~\eqref{eq:simplificationalpha} when
$\alpha = 0$ is:
\begin{equation}
  \label{eq:cdispersion}
  \bar{c}
  = \del{1 -
  \frac{2}{3}\del{\frac{120u_{B}}{\Ray}}^{\frac{1}{3}}x}^{\frac{3}{2}}.
\end{equation}
Substitution of this solution into equation~\eqref{eq:fluctuationsinc}
and comparison of terms identifies that the dominant balance is indeed
given by~\eqref{eq:balance}, in the limit $u_{B}^{2} \Ray \ll 1$
and~$u_{B} < 1$. The flow extends a large distance upstream compared
to the thickness of the aquifer and the cross-flow diffusion is fast
compared to the time for the background flow to pass through the
region in which there is an elevated CO$_2$ concentration. The
solution~\eqref{eq:cdispersion} suggests that the region of enhanced
concentration advances upstream a non-dimensional distance
\begin{equation}
  \label{eq:xdis}
  X_{\textrm{dis}}
  = \frac{3}{2}\del{\frac{\Ray}{120u_{B}}}^{\frac{1}{3}}.
\end{equation}

In the limit $\alpha \gg 1$ the steady-state is dominated by a balance
of advection and diffusion, and the solution may be approximated by
\begin{equation}
  \label{eq:cdiffusion}
  \bar{c} =  e^{-\Ray u_{B} x},
\end{equation}
with a characteristic length scale of
\begin{equation}
  \label{eq:xdif}
  X_{\textrm{diff}} = - \frac{\ln c_{\text{diff}}}{\Ray  u_{B}},
\end{equation}
where $c_{\text{diff}}$ is the concentration at which we consider it
to be negligible.

\begin{figure}
  \centering
  \includegraphics[width=0.7\textwidth]{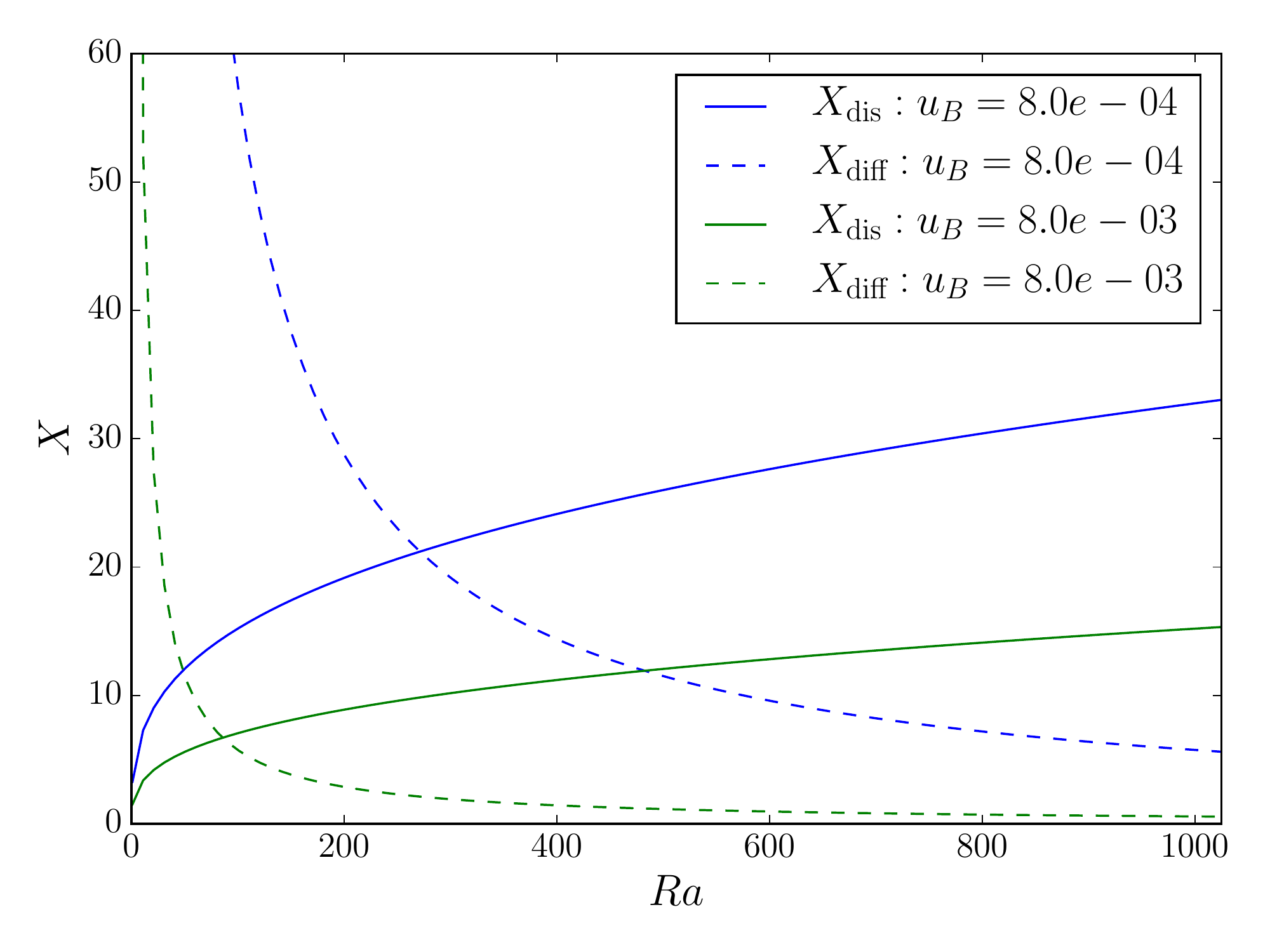}
  \caption{The distance the CO$_{2}$ front extends upstream in the
    dispersion limit and in the diffusion limit as a function of
    $\Ray$ for two different values of background flow~$u_{B}$.
    Distances $X$ correspond to multiples of the aquifer height~$H$.}
  \label{fig:xs_ra}
\end{figure}

Equating $X_{\text{diff}}$ and $X_{\text{dis}}$, we find that
\begin{equation}
  \alpha_{e} = \frac{3}{2 \ln c_{\text{diff}}},
\end{equation}
and this provides an indication of the transition between diffusive
and dispersive mechanisms. For $c_{\text{diff}} = 0.01$, we
find~$\alpha_{e} = 0.326$.  Figure~\ref{fig:xs_ra} shows
$X_{\text{dis}}$ and $X_{\text{diff}}$ as function of $\Ray$ for two
different values of the background flow, using~$c_{\text{diff}} =
0.01$.  For larger Rayleigh numbers, dispersion controls the distance
that the CO$_{2}$ front extends upstream.

\subsection{Regime differentiation}

\begin{figure}
  \centering
  \includegraphics[width=0.7\textwidth]{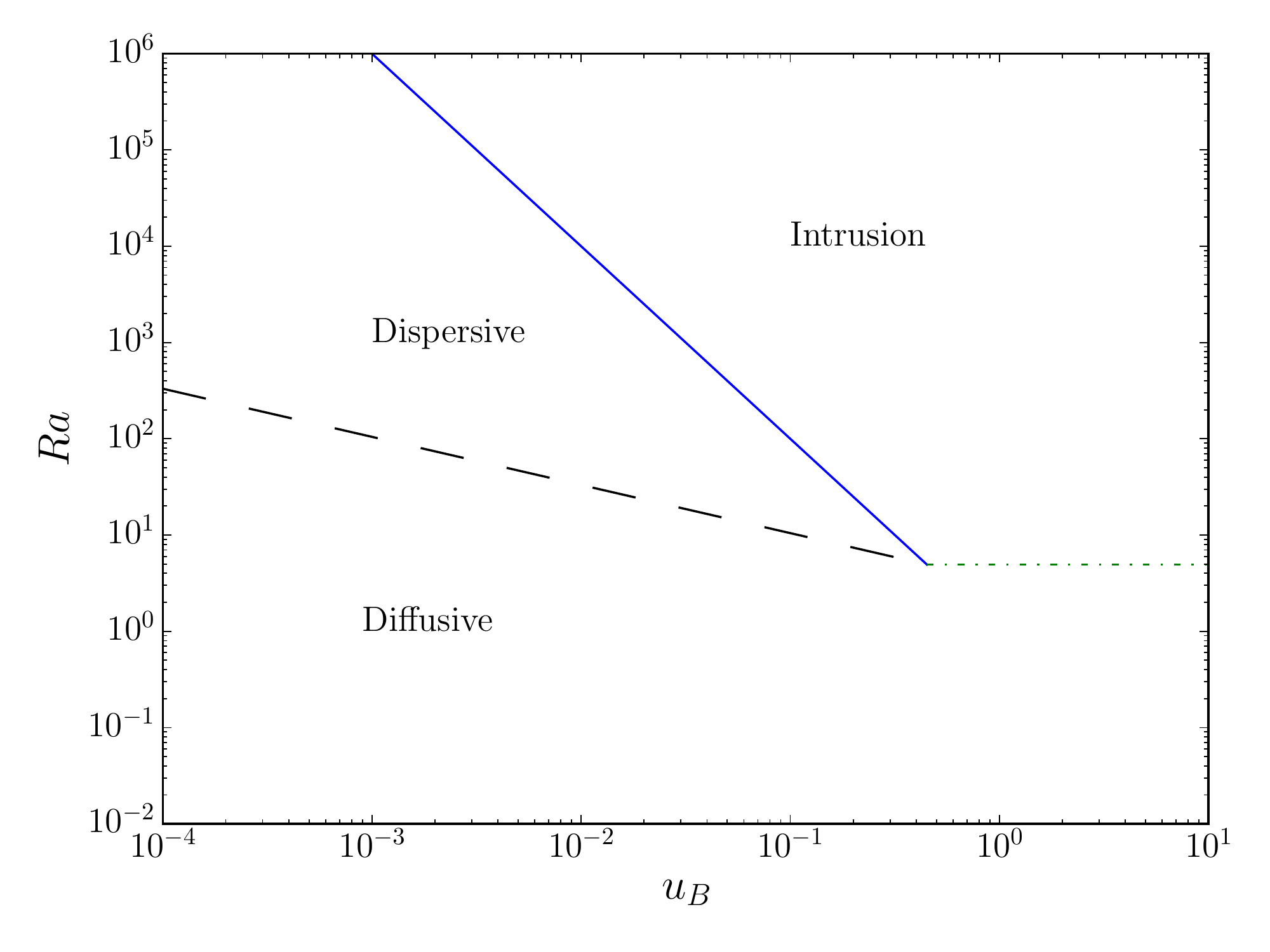}
  \caption{Illustration of the different regimes. The dashed line
    denotes $\alpha = 1$ (see equation~\eqref{eq:alpha}) which
 	delineates the  dispersive and diffusive regimes for $u_B < O(1)$.
    The solid line shows $\Ray = 1/u_{B}^{2}$ (see
    equation~\eqref{eq:intrusion}) and delineates the boundary between
    the intrusion regime and (i) the dispersive regime for $u_B<O(1)$
    and (ii) the diffusive  regime for $u_{B}>O(1)$. The dotted line
    $\Ray= 4.93$ denotes an upper bound on the transition between the
    intrusive and diffusive regimes in the case $u_{B}>1$, when the
    intrusion is relatively short, and the along-aquifer diffusive
    transport dominates the cross-aquifer diffusion; we have chosen
    $\Ray = 4.93$, so that this line intersects the point where the
    dashed and solid lines converge.}
  \label{fig:regimecomparison}
\end{figure}

Combining the analysis of the gravity intrusion with the model of the
buoyancy-driven shear dispersion, we infer that for $u_{B} < O(1)$ three
regimes may arise, as shown in Figure~\ref{fig:regimecomparison}.
Gravity intrusion occurs when $u_{B}^{2} \Ray \gg 1$. When $u_{B}^{2}
\Ray \ll 1$, either a diffusion or dispersion dominated flow results,
depending on the parameter $\alpha$ (see equation~\eqref{eq:alpha}).

As $u_{B}$ approaches unity, the above analysis shows that the
transition between the dispersion and diffusion regimes and between
the dispersion and intrusion regimes converge. In the case $u_{B}
\geq O(1)$, the flow becomes more restricted in lateral extent upstream
of the anticline, and our analysis of time-scales for $u_{B}>1$ given at
the end of Section~\ref{sec:gravity_intrusion}, suggests that with
$u_{B}>O(1)$, an upper  bound for the case in which the along-aquifer
diffusion dominates the intrusion regime is  $ \Ray = 0(1)$.  In the
diffusion dominated regime, from the boundary
conditions~\eqref{eq:cbc1And} and~\eqref{eq:dbc1and}, we expect
that at $x=0$, $\bar{c} < 1$ and we explore this further in
Section~\ref{sec:stong_bg_flow} below. In Figure~\ref{fig:regimecomparison},
we illustrate this upper bound with a dotted line, which for convenience we
show as $\Ray = 4.93$ so that it intersects the point at which the solid
and dashed lines converge.

\section{Comparison of analytical and numerical models}
\label{sec:examples}

To support the asymptotic analysis, the full problem in
Section~\ref{sec:full_model} has been solved on a domain of length $L
= 100$ and height~$H = 1$.  We use a mixed finite element method for
the Darcy flow, and an upwinded discontinuous Galerkin method for the
transport equation. Problems are advanced in time until a steady-state
is reached.  A detailed description of the numerical method and the
complete computer code used to produce all examples is provided in the
supporting material~\citep{unwin:supporting}.  The computer code is
built on the FEniCS libraries~\citep{fenics:book}.

\subsection{Weak background flows ($u_{B} < 1 $)}

\begin{figure}
  \begin{subfigure}[b]{\textwidth}
    \center
    \vspace{1.0em}
    \includegraphics[width=0.6\textwidth]{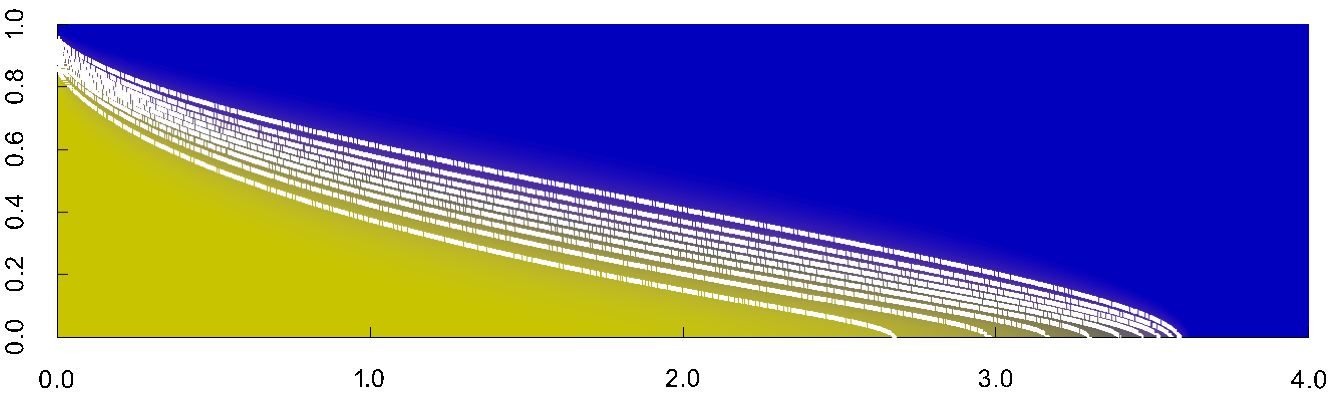}
    \caption{Gravity intrusion: $\Ray = 3000 $}
    \label{fig:grav_int}
  \end{subfigure}
  \begin{subfigure}[c]{\textwidth}
    \center
    \vspace{1.0em}
    \includegraphics[width=0.6\textwidth]{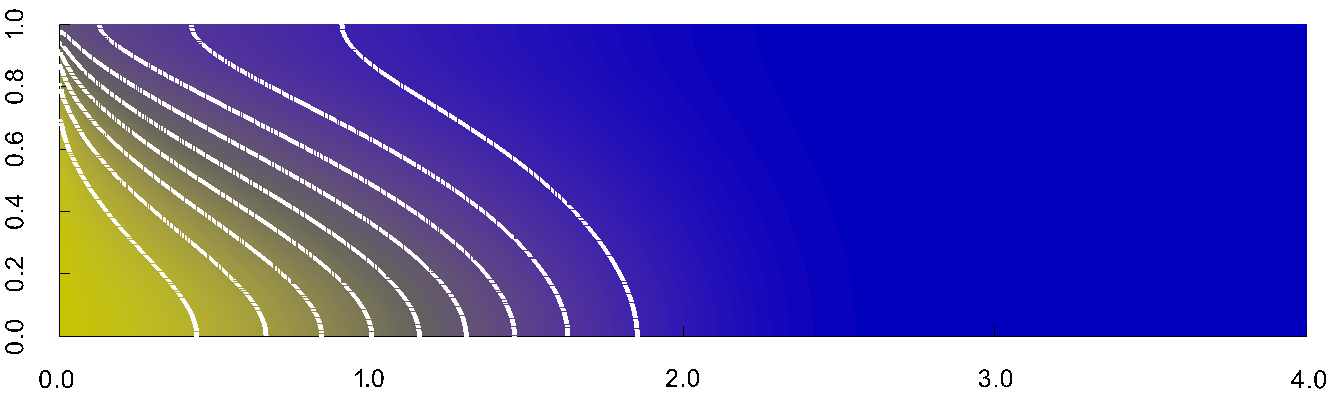}
    \caption{Dispersive regime: $\Ray = 50$}
  \end{subfigure}
  \begin{subfigure}[c]{\textwidth}
    \center
    \vspace{1.0em}
    \includegraphics[width=0.6\textwidth]{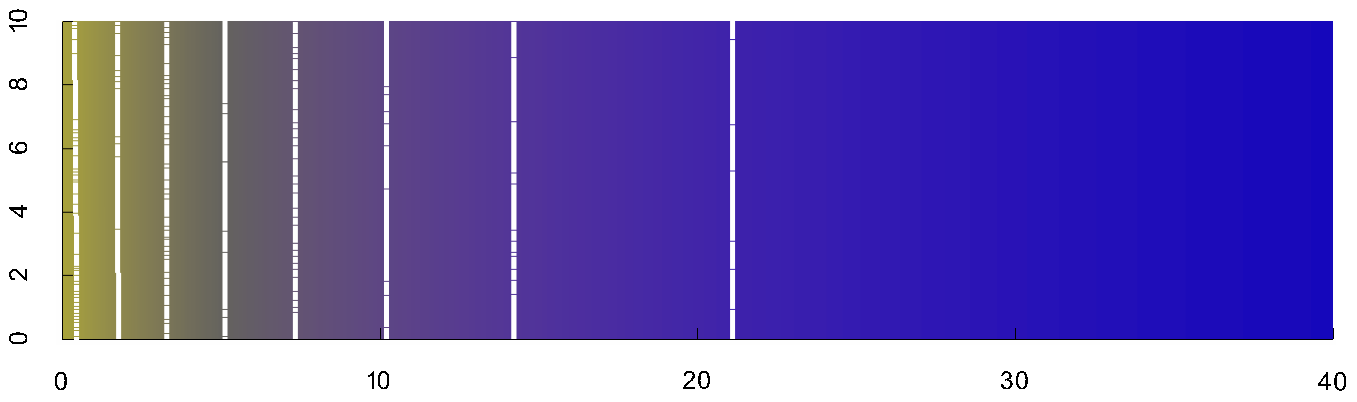}
    \caption{Diffusive regime: $\Ray = 1.0$}
  \end{subfigure}
  \begin{subfigure}[b]{\textwidth}
    \center
    \includegraphics[width=0.3\textwidth]{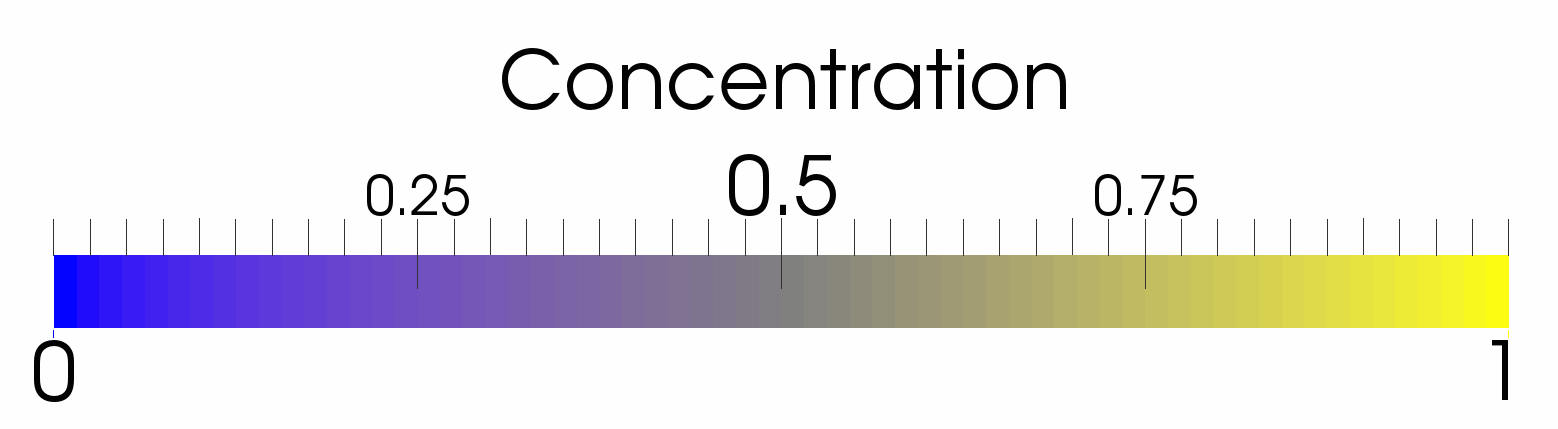}
  \end{subfigure}
  \caption{Concentration contours for $u_{B} = 0.1$ and different
    values of~$\Ray$.  The colours show the concentration locally as
    defined by the scale in the figure and the contour lines are shown
    at equal intervals of 0.1 from 0.1 to 0.9. The $x$-axis is longer
    in (c) than (a) or (b) to display the full diffusive regime.}
  \label{fig:flowplots}
\end{figure}

To illustrate the form of the velocity and concentration fields for
the three different regimes when $u_{B} < 1$, we show in
Figure~\ref{fig:flowplots} the computed concentration field for three
different values of $\Ray$ when~$u_{B} = 0.1$. The values of $\Ray$
correspond to points in the gravity intrusion, dispersion dominated,
and diffusion dominated regimes.

\subsubsection{Concentration profiles}
\label{sec:concprofile}

\begin{figure}
  \centering
  \includegraphics[width=0.7\textwidth]{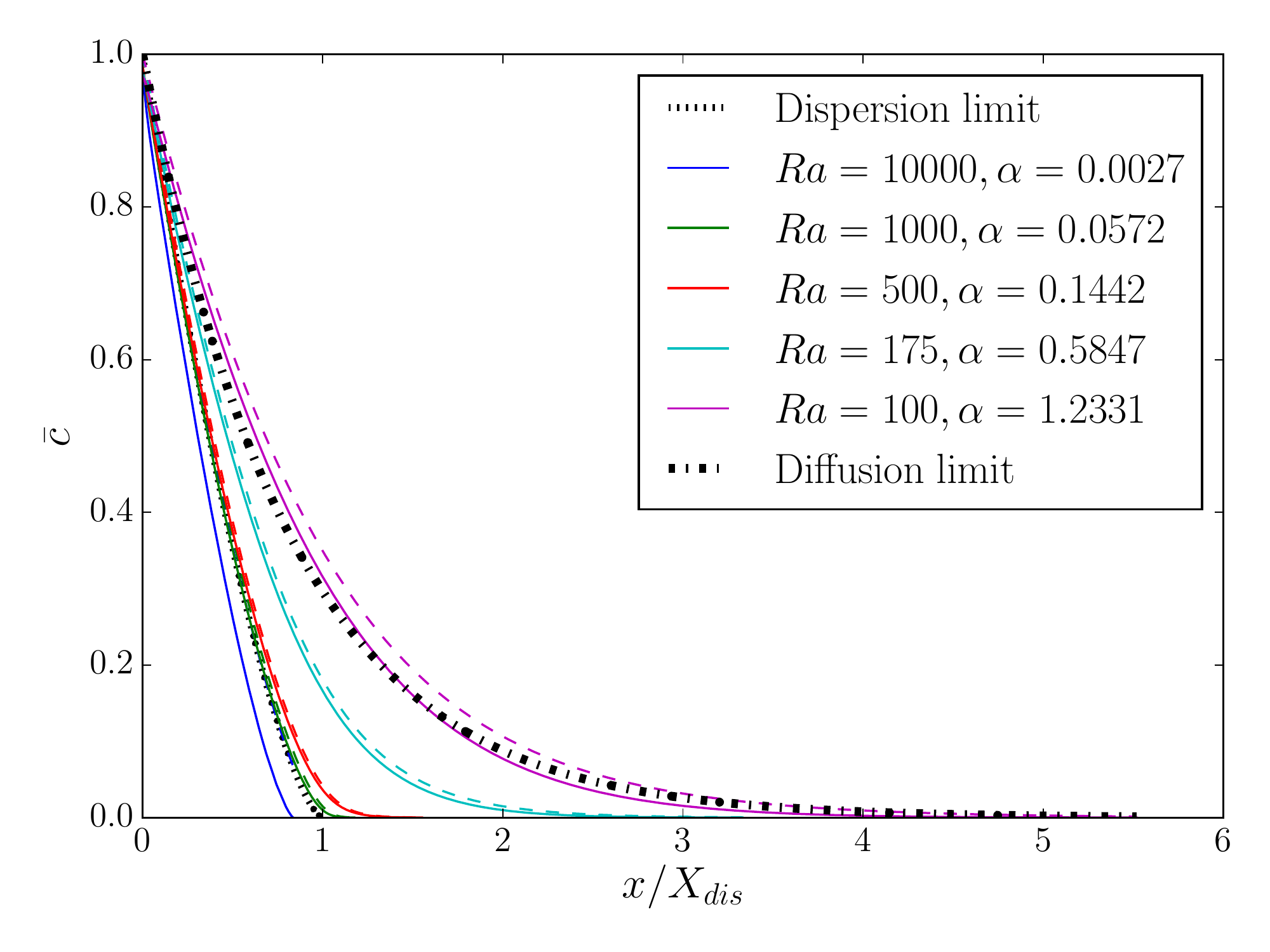}
  \caption{Computed and analytical variation of $\Bar{c}$ with $x$ for
    different values of~$\Ray$ when $u_{B} = 8 \times 10^{-4}$.  The
    solid lines represents the two-dimensional numerical solution and
    the dashed lines represent the one-dimensional solution.}
    \label{fig:conc_profiles}
\end{figure}

Figure~\ref{fig:conc_profiles} shows the vertically averaged
concentration as a function of position along the aquifer for the
one-dimensional model as given by ~\eqref{eq:simplificationalpha} (dashed lines)
and the  two-dimensional full numerical calculations (solid lines) for $u_{B} = 8 \times
10^{-4}$ and values of $\Ray$ ranging from $100$ to~$10000$.
The one-dimensional model (equation~\eqref{eq:simplificationalpha})
was solved using the FEniCS libraries and the full code is included in
the supporting material~\citep{unwin:supporting}.  There is very good
agreement between the two models when $\Ray \le 1000$ and the flow is
in either the dispersion or diffusion dominated regime. The limiting
dispersive~\eqref{eq:cdispersion} and diffusive~\eqref{eq:cdiffusion}
cases (heavy black lines as shown in legend) are also shown in the figure, and coincide with
the two-dimensional numerical simulation when in the appropriate limit
regimes.

When $\Ray = 10000$, the problem is entering the gravity intrusion
regime. In Figure~\ref{fig:conc_profiles} the two-dimensional
numerical solution for this case no longer matches the one-dimensional
dispersion dominated model in~\eqref{eq:simplificationalpha} since
there are significant fluctuations in concentration across the height
of the aquifer.  In this regime   the solution has a
narrow vertical region of adjustment in the concentration field from
the CO$_2$ saturated fluid at the base of the aquifer to the
unsaturated oncoming groundwater at the top of the aquifer,
reminiscent of the intrusion model given in
Section~\ref{sec:gravity_intrusion}.

\begin{figure}
  \center
  \includegraphics[width=0.7\textwidth]{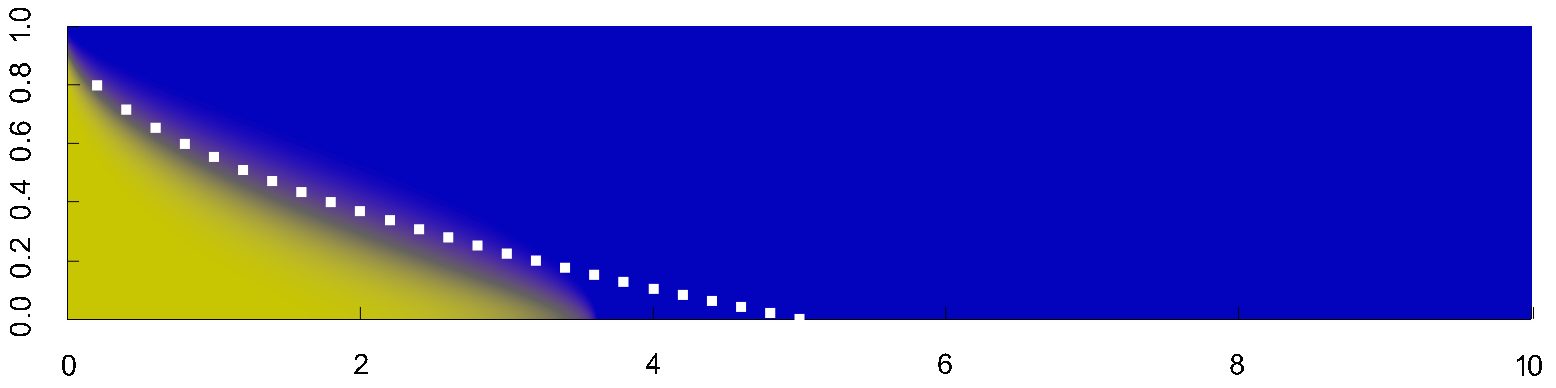}
  \caption{Concentration as a function of position  for the gravity intrusion case with
    $u_{B} = 0.1$ and $\Ray = 3000$  and the prediction of the  intrusion model
    (equation~\ref{eq:intrusion_height}) for the interface position
    overlaid (dotted line).}
  \label{fig:intrusion}
\end{figure}

Indeed, in
Figure~\ref{fig:intrusion} we compare the numerical solution for the
concentration in the case $\Ray=3000$ and $u_B=0.1$ with the prediction of the interface height $h$
as predicted by equation~\eqref{eq:intrusion_height} for the gravity
intrusion model. That model
treats the adjustment of the concentration from the CO$_2$ saturated
fluid to the oncoming groundwater flow, as   a sharp
interface.  There is a reasonable match for $x < 3.5$.  However, the
diffusive boundary layer which is present in the full numerical solution leads
to a weak recirculation that is not included in the intrusion model.

\subsubsection{Velocity fluctuation profiles}

\begin{figure}
  \centering
  \begin{subfigure}[b]{0.7\textwidth}
    \centering \includegraphics[width=\textwidth]{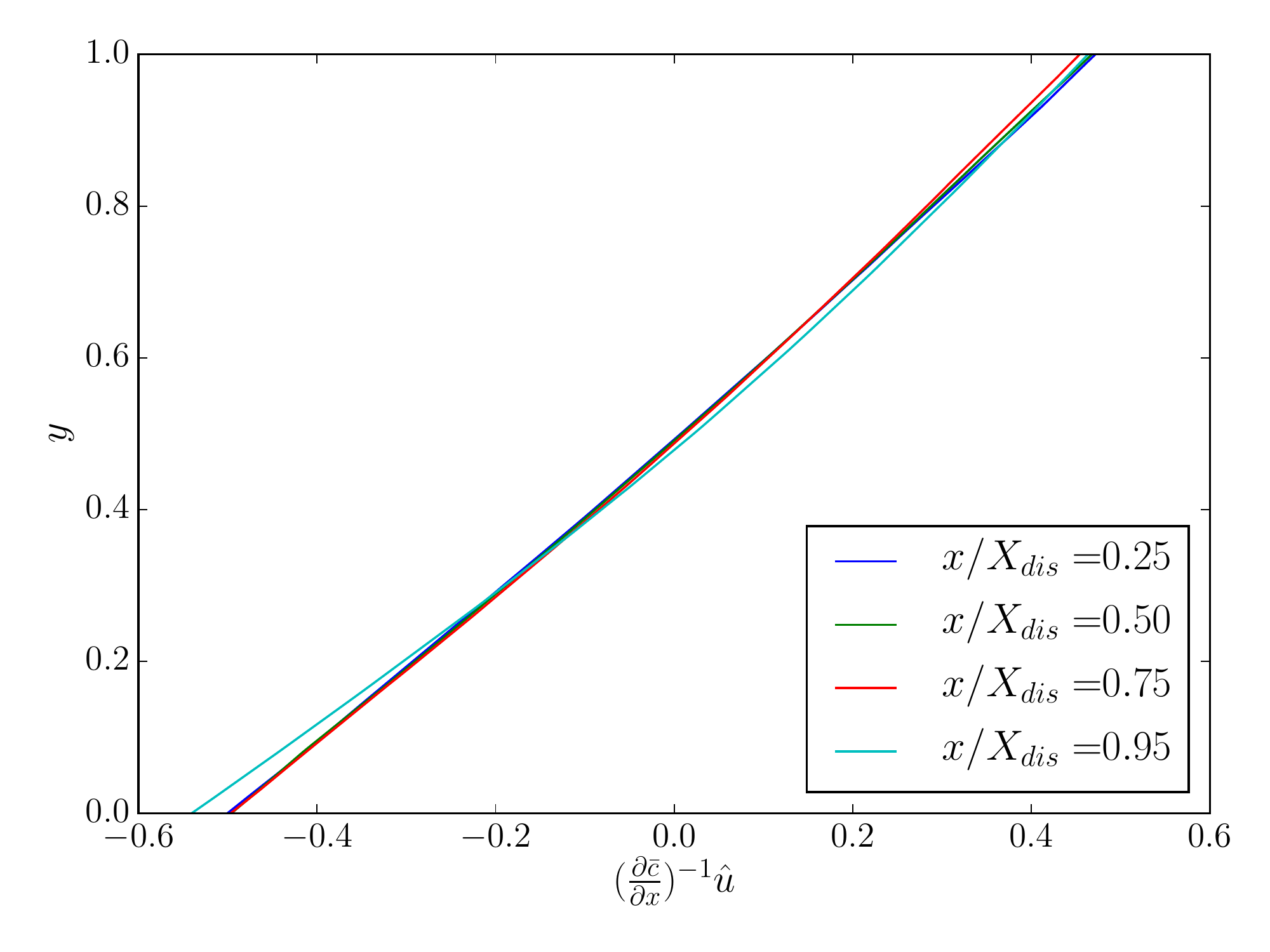}
    \caption{Velocity fluctuations $\hat{u}$ scaled by concentration
      gradient for $\Ray = 1000$ and $u_{B} = 8 \times 10^{-4}$
      ($\alpha = 0.0527$).}
    \label{fig:scaled_velocity_profile}
  \end{subfigure}
  \\
  \begin{subfigure}[b]{0.7\textwidth}
    \centering
    \includegraphics[width=\textwidth]{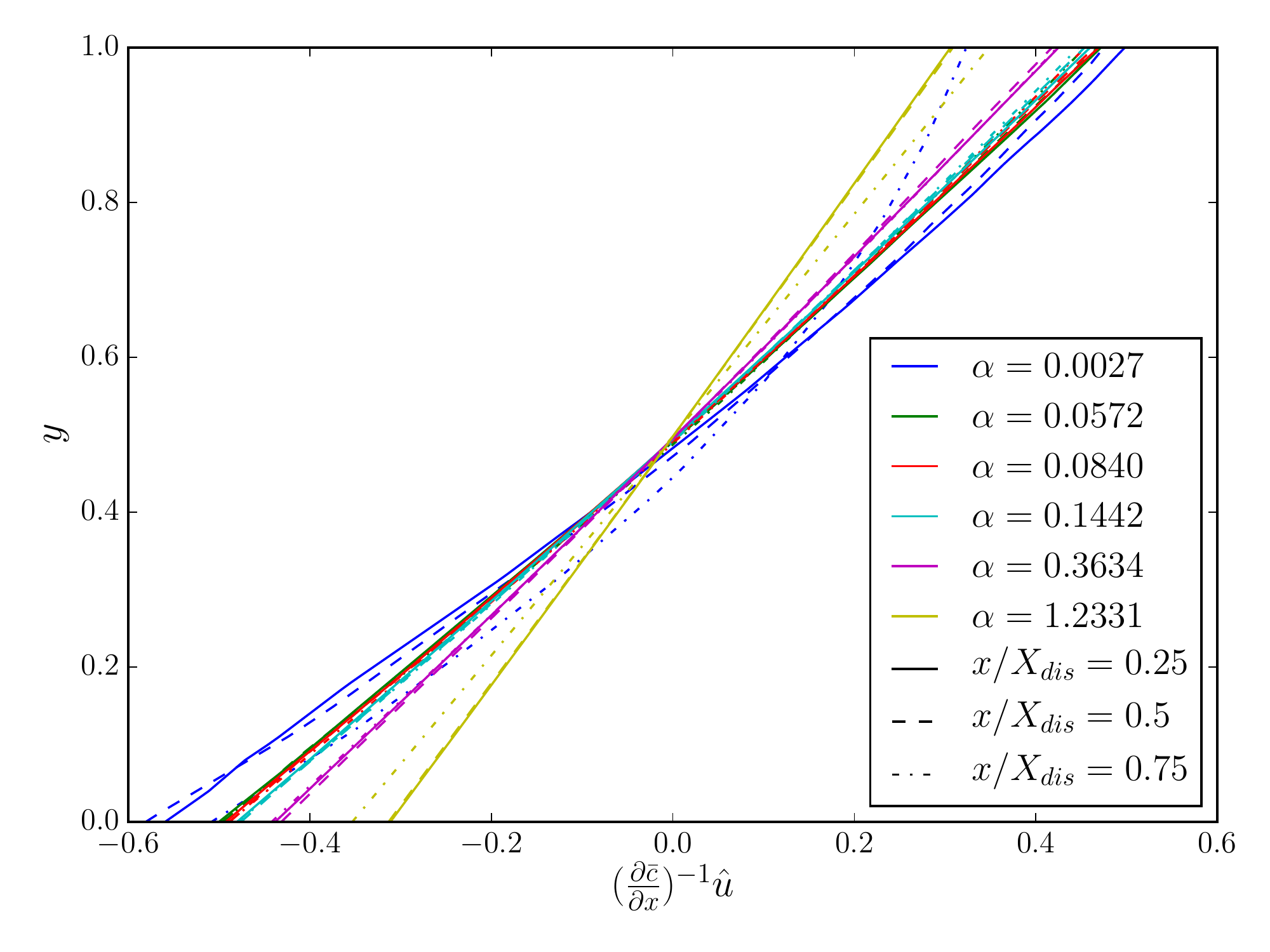}
    \caption{Velocity fluctuations $\hat{u}$ for different values of
      $\alpha$ at different values of~$x$.}
    \label{fig:v_alpha}
  \end{subfigure}
  \caption{Scaled velocity profiles at different points along the
    domain for different values of~$\alpha$ with $u_B=8 \times 10^{-4}$.}
  \label{fig:v_profiles}
\end{figure}

In the buoyancy driven dispersion regime, which arises for  small $\alpha$ and when $u_{B}^{2} \Ray \ll 1$,
equation~\eqref{eq:uflucprof} predicts that the velocity variation
from the mean flow $\hat{u}$ will vary linearly with depth.
If~$\hat{u}$ is divided by the depth-averaged concentration gradient,
the profiles are predicted to pass through $-0.5$ at $y = 0$ and $0.5$
at~$y = 1$.  Figure~\ref{fig:scaled_velocity_profile} shows the scaled
velocity profiles computed from the two-dimensional model for $\Ray =
1000$ and $u_{B} = 8 \times 10^{-4}$ ($\alpha = 0.0527$).  For $x
\lesssim (3/4)X_{dis}$ the velocity profiles are linear, whereas close
to the stall point ($x/X_{dis} = 1$) the profile begins to deviate
from the simplified theory.
\begin{figure}
    \centering
    \includegraphics[width=0.7\textwidth]{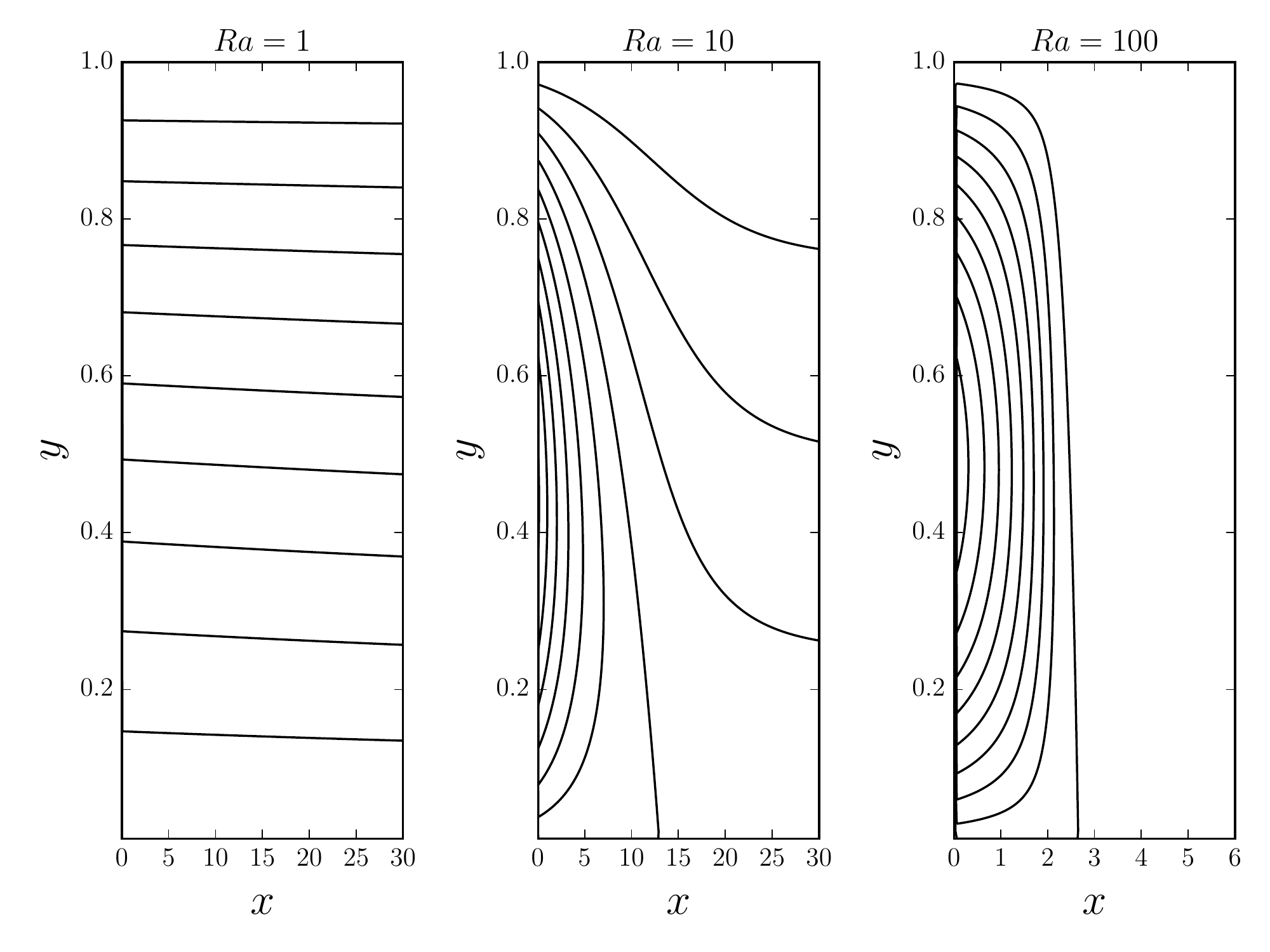}
    \caption{Typical streamlines of the flow for $u_{B} = 0.01$ and
    $\Ray = 1, 10$ and $100$.}
    \label{fig:streamfunctions}
\end{figure}

When in the regime where diffusion is important ($u_{B}^{2} \Ray \ll
1$), as $\alpha$ increases and the flow transitions from dispersive to
diffusive, the one-dimensional analytical model becomes less
applicable for~$\hat{u}$.  This can be seen in
Figure~\ref{fig:v_alpha}, where the scaled horizontal velocity
fluctuations are shown at different distances into the domain for
different values of~$\alpha$.  For $0.0572 \leq \alpha \leq 0.1442$,
the scaled horizontal velocity fluctuations at different distances
into the domain all lie on a straight line between $-0.5$ and~$0.5$.
As $\alpha$ increases, the numerically computed profiles deviate from
the model, as the lateral diffusive transport becomes comparable to
and then progressively more significant than the shear dispersion, but in this diffusion limit
 these perturbation velocities are very small compared to the background
hydrological flow.
When $\Ray = 10000$ and $u_{B} = 8 \times 10^{-4}$ ($\alpha = 0.0027$)
the gravity intrusion region is being approached and the vertical concentration
fluctuations across the domain are no longer small and so the velocity
fluctuations are no longer governed by
equation~\eqref{eq:uflucprof}.

In Figure~\ref{fig:streamfunctions}, we present a picture of the
typical streamlines for the flow when $u_B = 0.01$ and $\Ray = 1, 10$
and $100$. The streamlines are computed, approximately, by solving
equation~\eqref{eq:simplificationalpha} numerically and inserting the
result into~\eqref{eq:uflucprof}, and then adding on the background
hydrological flow.  The case $\Ray = 1$ corresponds to
the diffusion limit, the case $\Ray = 100$ corresponds to the
dispersion regime, and the case $\Ray = 10$ is an intermediate
case. The figure shows that in the diffusion limit the oncoming
groundwater flow is dominant and the flow remains nearly uniform, but as $\Ray$ increases, some
fluid begins to flow upstream leading to a small circulation in the
lower part of the aquifer. As $\Ray$ increases further, and the flow
is controlled by the buoyancy driven dispersion, a strong
recirculation develops upstream of the anticline, leading to the
diversion of the oncoming groundwater flow towards the top of the
aquifer.

\subsection{Strong background flows ($u_{B} \gtrsim 1 $)}
\label{sec:stong_bg_flow}

\begin{figure}
  \begin{subfigure}[b]{\textwidth}
    \center
    \includegraphics[width=0.6\textwidth]{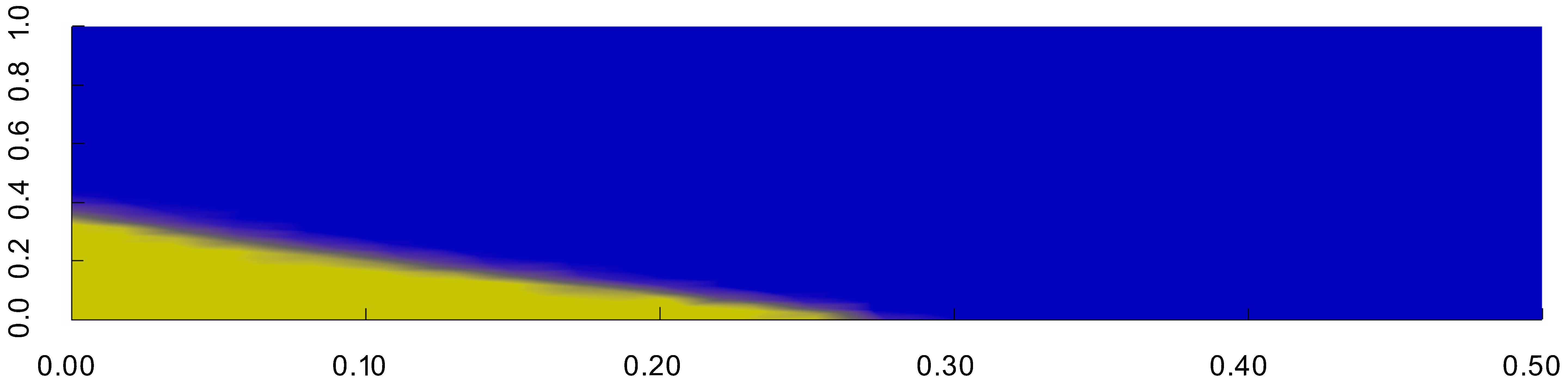}
    \caption{Gravity intrusion regime: $\Ray = 1000$}
    \label{fig:flowplots2-intrusion}
  \end{subfigure}
  \begin{subfigure}[c]{\textwidth}
    \center
    \vspace{1.0em}
    \includegraphics[width=0.6\textwidth]{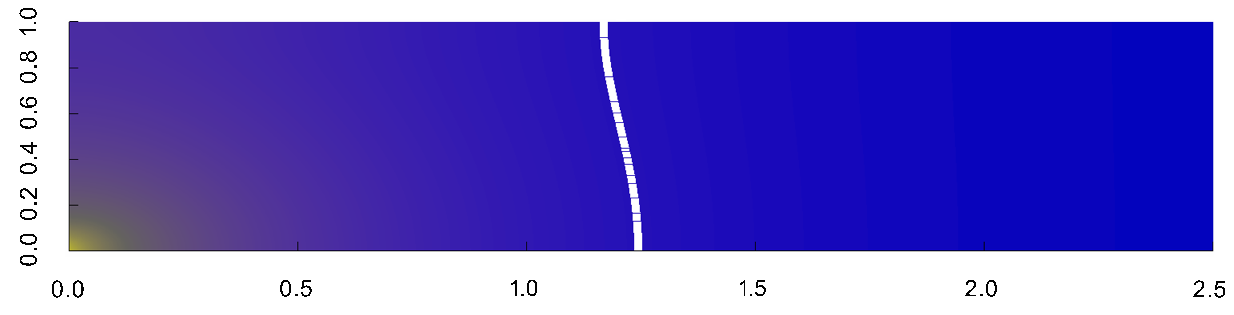}
    \caption{Transitional regime: $\Ray = 2 $}
    \label{fig:flowplots2-transition}
  \end{subfigure}
  \begin{subfigure}[c]{\textwidth}
    \center
    \vspace{1.0em}
    \includegraphics[width=0.6\textwidth]{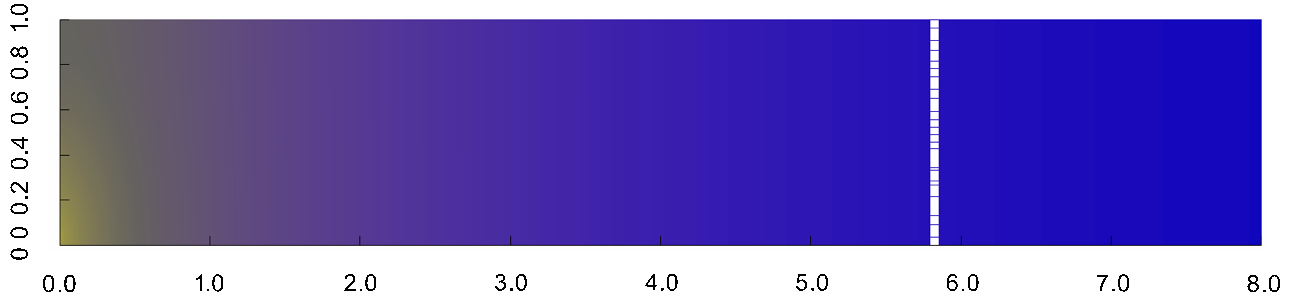}
    \caption{Diffusive regime: $\Ray = 0.5$}
    \label{fig:flowplots2-diffusive}
  \end{subfigure}
  \begin{subfigure}[b]{\textwidth}
    \center
    \vspace{1.0em}
    \includegraphics[width=0.3\textwidth]{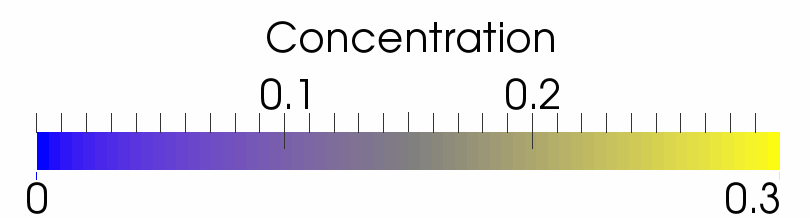}
  \end{subfigure}
  \caption{Concentration fields for different values of~$\Ray$ for
    $u_{B} = 1.0$. The white contour in (b) and (c) is the~$c = 0.01$
    contour. The presented domains have been truncated at different
    lengths to best show each regime.  Note that the maximum
    concentration for the colour bar is set equal to 0.3, which is
    greater than the highest concentration for panels (b) and~(c). }
  \label{fig:flowplots2}
\end{figure}

We now look at the case with a stronger  background flow  ($u_{B} \gtrsim1$). Three examples of full
numerical solutions are shown in Figure~\ref{fig:flowplots2}
corresponding to $\Ray = 1000$, $2$ and $0.5$ for~$u_{B}=1.0$.
In contrast to the case $u_{B} < O(1)$, only two distinct regimes
develop (figure 3) .  Now, the  flow transitions from the gravity intrusion regime to the
diffusion dominated adjustment of the concentration since, with large $u_B$,  the upstream extent of the buoyancy driven flow
is insufficient for the buoyancy driven dispersion to develop before the along aquifer diffusion becomes significant.  With large
$\Ray$, the solution is similar to the gravity-driven intrusion
solution (Figure~\ref{fig:flowplots2-intrusion}), while for smaller
values of $\Ray$ the solution evolves towards the along aquifer
diffusion solution (Figure~\ref{fig:flowplots2-diffusive}).

\begin{figure}
  \centering
  \includegraphics[width=0.7\textwidth]{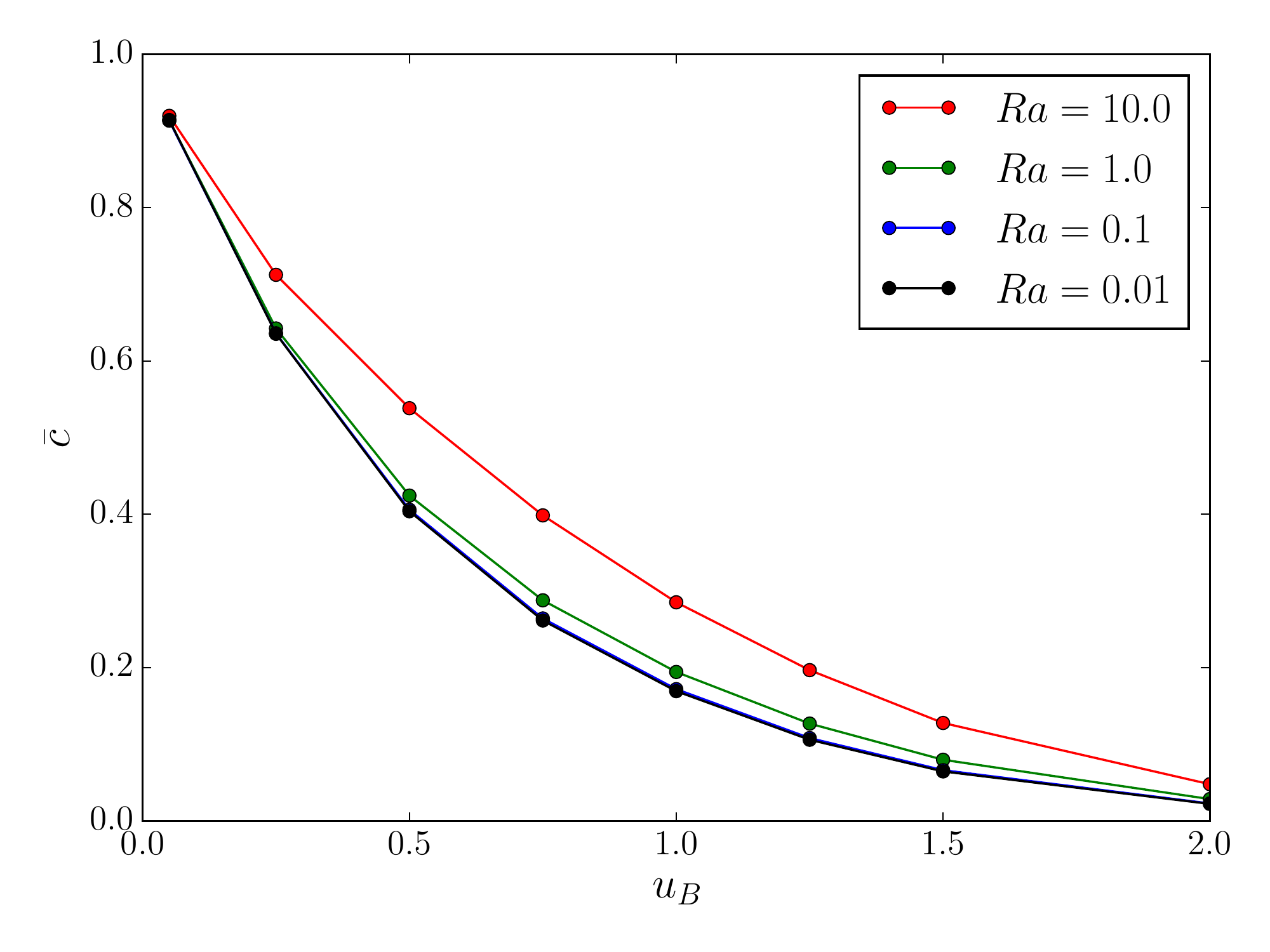}
  \caption{Computed vertically averaged concentration for various
    values of $\Ray$ and $u_{B}$ at~$x = 0$.}
  \label{fig:cbarx0}
\end{figure}

We have calculated the vertically
averaged mean concentration at $x = 0$ from the numerical solutions
for the cases when $\Ray = 10, 1, 0.1$ and $0.01$, as shown in
Figure~\ref{fig:cbarx0}.  As $u_{B}$ increases, $\bar{c}$ decreases at
$x = 0$ since there is a progressively stronger flow from the upstream
region which suppresses the upstream buoyancy driven flow of dense,
CO$_2$ saturated fluid from below the anticline

\begin{figure}
  \centering
  \includegraphics[width=0.7\textwidth]{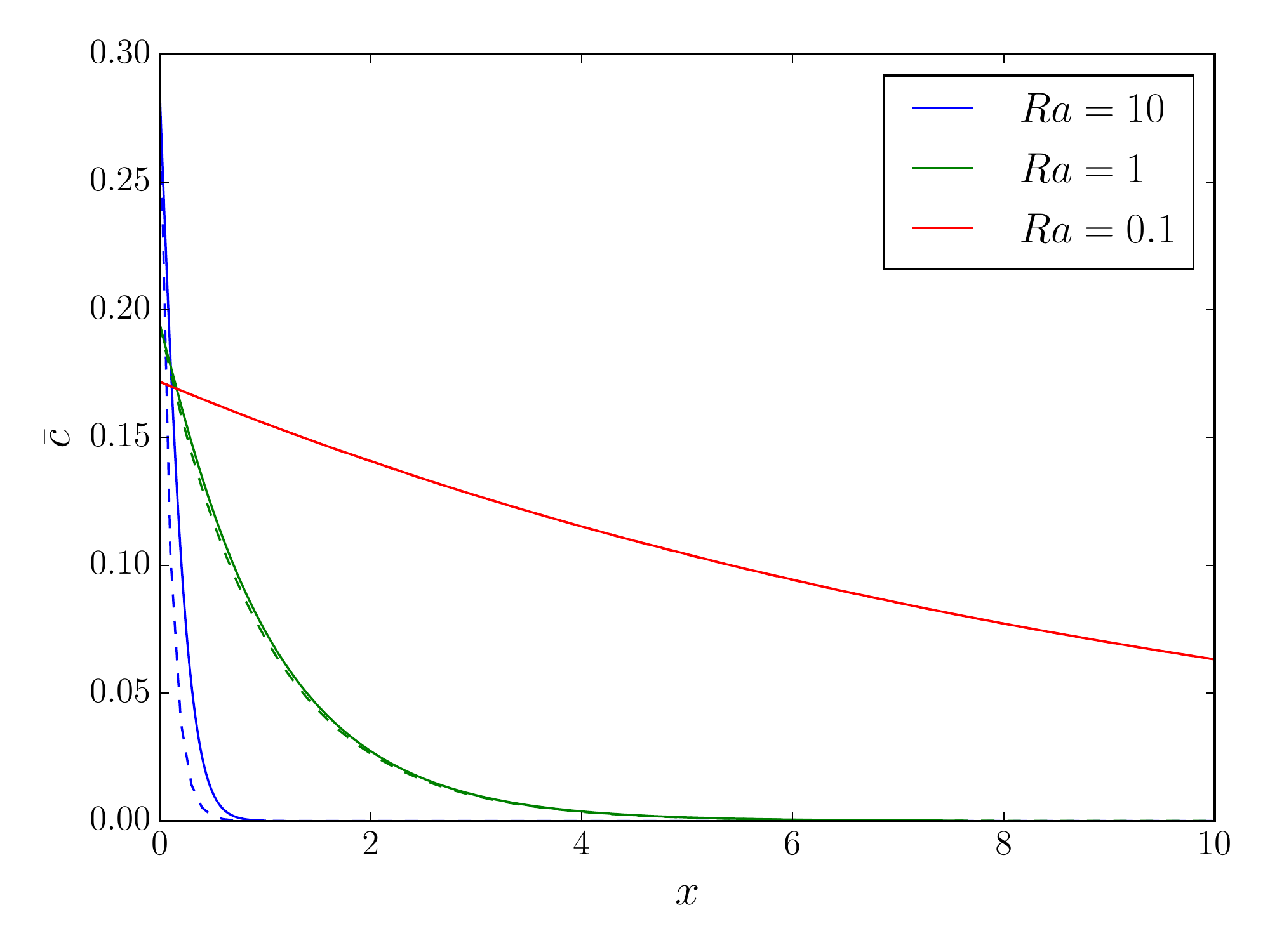}
  \caption{Computed and analytical variation of $\Bar{c}$ with $x$ for
    different values of~$\Ray$ when~$u_{B} = 1.0$.  The solid line
    represents the two-dimensional numerical solution and the dashed
    line represents the diffusion limit if using the numerically
    computed $\Bar{c}$ as the boundary condition at~$x = 0$.}
  \label{fig:conc_profile_diffusion}
\end{figure}

In the diffusive regime, which we expect to apply for small  $\Ray   $,
 the vertically averaged concentration may be approximated
by the diffusive solution of equation~\eqref{eq:simplificationalpha},
given by:
\begin{equation}
  \bar{c}(x) = \bar{c}(0) e^{-\Ray u_{B} x}.
  \label{eq:cbar_c0}
\end{equation}
Figure~\ref{fig:conc_profile_diffusion} shows the vertically-averaged
concentration profiles for the three values of $\Ray$ and $u_{B} =
1.0$. Using the numerically determined value for the mean
concentration at $x=0$, we have compared the vertically averaged
concentration with the diffusion solution given by
equation~\eqref{eq:cbar_c0}. When $\Ray = 0.1$, the system is in the
diffusion regime and the numerical solution for $\Bar{c}$ matches the
diffusion profile. When $Ra \geq 1$, the simulations move towards the
intrusion regime and the numerical solutions evolve away from the
approximate analytical solution.

\section{Conclusions}

We have explored both analytically and numerically the long-term
dissolution of a plume of CO$_2$ trapped in an anticline and driven by
a steady background flow of CO$_2$ unsaturated water from upstream. We
have focused on the role of diffusion and buoyancy-driven flow in
regulating the distribution of CO$_2$ in the aquifer fluid upstream of
the anticline. In the case $u_{B} < 1$, where $u_{B}$ is the
dimensionless background hydrological flow, the buoyancy-driven speed
of the dense CO$_2$ saturated water exceeds the oncoming flow speed
and the CO$_2$ extends a significant distance upstream of the
anticline ($x \gg H$).  In this case, we have established that three
different regimes may develop. With a small diffusive flux across the
aquifer ($u_{B}^{2} \Ray \gg 1$, where $\Ray$ is the Rayleigh number
associated with the dense CO$_2$ laden fluid), a static intrusion of
dense CO$_2$ saturated fluid develops and extends a distance
$1/2u_{B}$ upstream of the anticline. This is balanced by the pressure
gradient in the oncoming flow.  With larger diffusive fluxes across
the aquifer ($u_{B}^{2} \Ray \ll 1$) we have established that a
buoyancy-driven shear dispersion flow regime may develop and a
convective recirculation develops just upstream of the aquifer,
regulated by (i)~the supply of unsaturated aquifer fluid from
upstream; (ii)~the buoyancy-driven flow associated with the dense
CO$_2$ saturated fluid from downstream; and (iii)~the vertical
diffusion of CO$_2$ across the aquifer.  However, if the diffusive
transport is too rapid ($\alpha \ll 1$) then a simple
advection-diffusion balance regulates the distribution of the CO$_2$
in solution in the water upstream of the anticline.  In the case
$u_{B} > 1$, the CO$_2$ extends a much smaller distance upstream from
the anticline, and in this case, either only the advection-diffusion
balance or the intrusion regimes develop.

In the context of CO$_2$ sequestration in deep saline aquifers, this
analysis is important as it demonstrates the strong effect that a
background hydrological flow has on the long-term dissolution of
CO$_2$ in a structural trap.  We now show that under some typical
conditions the dynamics may indeed be controlled by a balance between
the buoyancy-driven shear dispersion and the background hydrological
flow.  This leads to new estimates of the maximum upstream migration
of CO$_2$ rich groundwater.  The solubility of CO$_2$ in groundwater
is only a few wt\%, so if we consider a plume of CO$_2$ of order 10~m
deep, trapped in a structural anticline and connected to a laterally
extensive aquifer of order 20--30~m deep, then vertical convective
dissolution alone will only lead to dissolution of order 0.2--0.6~m.
Continued dissolution will require the lateral supply of
undersaturated water from the aquifer and this may be achieved through
a combination of buoyancy-driven lateral dispersion of the dense
CO$_2$ saturated water from below the CO$_2$ plume and supply of water
resulting from a background hydrological flow. For typical conditions,
with permeability of order $0.1$--$0.01$ Darcy, a density difference
of order a few percent between the undersaturated and saturated water,
and an aquifer diffusivity of order $10^{-9}$--$10^{-10}$~m$^{2}$/s,
the Rayleigh number will be of order~$10^{3}$--$10^{4}$. With a
hydrological flow speed of order $10^{-8}$--$10^{-9}$~m/s, the
dimensionless velocity $u_{B}$ will be of order~$0.01$--$1.0$.  From
Figure~\ref{fig:regimecomparison} we see that it is the transport
associated with the shear dispersion that balances the steady
background flow.  We estimate that the length scale of the dispersive
transport (equation~\eqref{eq:xdis}) will be of order $100$--$400$~m.
Once the steady flow regime is established, the continued dissolution
will occur at a rate proportional to the supply of undersaturated
water in the hydrological flow, as given in non-dimensional form by
$u_{B}(c_{D} - c_{0})$. For an anticline whose extent in the direction
of the flow is of order $1000$~m, and with a CO$_2$ plume with initial
depth of order $10$~m, then in order to dissolve, this will require a
net flow of groundwater of order $10^{6}$~m$^{2}$, which will require
a time of order $10^{12}$--$10^{13}$~s corresponding to
$10^{5}$--$10^{6}$ years.

\subsubsection*{Acknowledgements}

HJTU was funded by an EPSRC Doctoral Training Partnership scheme
(grant EP/J500380/1). Data relating to this publication is available
in~\citet{unwin:supporting}.

\bibliographystyle{abbrvnat}
\bibliography{bibliography}
\end{document}